# Eugene Garfield and Algorithmic Historiography:

# Co-Words, Co-Authors, and Journal Names

Loet Leydesdorff

University of Amsterdam, Amsterdam School of Communications Research (ASCoR),

Kloveniersburgwal 48, 1012 CX Amsterdam, The Netherlands;

loet@leydesdorff.net; http://www.leydesdorff.net

**Abstract**

Algorithmic historiography was proposed by Eugene Garfield in collaboration with Irving Sher in the 1960s, but further developed only recently into HistCite™ with Alexander Pudovkin. As in history writing, HistCite™ reconstructs by drawing intellectual lineages. In addition to cited references, however, documents can be attributed a multitude of other variables such as title words, keywords, journal names, author names, and even full texts. New developments in multidimensional scaling (MDS) enable us not only to visualize these patterns at each moment of time, but also to animate them over time. Using title words, co-authors, and journal names in Garfield's oeuvre, the method is demonstrated and further developed in this paper (and in the animation at http://www.leydesdorff.net/garfield/animation). The variety and substantive content of the animation enables us to write, visualize, and animate the author's intellectual history.

**Introduction**

In a lecture entitled "From Computational Linguistics to Algorithmic Historiography," Garfield (2001) provided an account of his gradual invention of "algorithmic historiography" which ultimately led to the program HistCite™ that was introduced with the paper in *JASIST* in 2003 coauthored with Pudovkin and Istomin (Garfield *et al*., 2003a). After the invention of "bibliographic coupling" by Kessler in 1963 (Kessler, 1963), Garfield, Sher, & Torpie (1964) developed the concept of the citation as a recursive operation in a network and used this to map a historical reconstruction of the development of DNA: from Mendel (1865) to Nirenberg & Matthaei (1961-1962). These authors claimed (at p. ii) that "The citation network technique does provide the scholar with a new *modus operandi* which, we believe, could and probably will significantly affect future historiography."

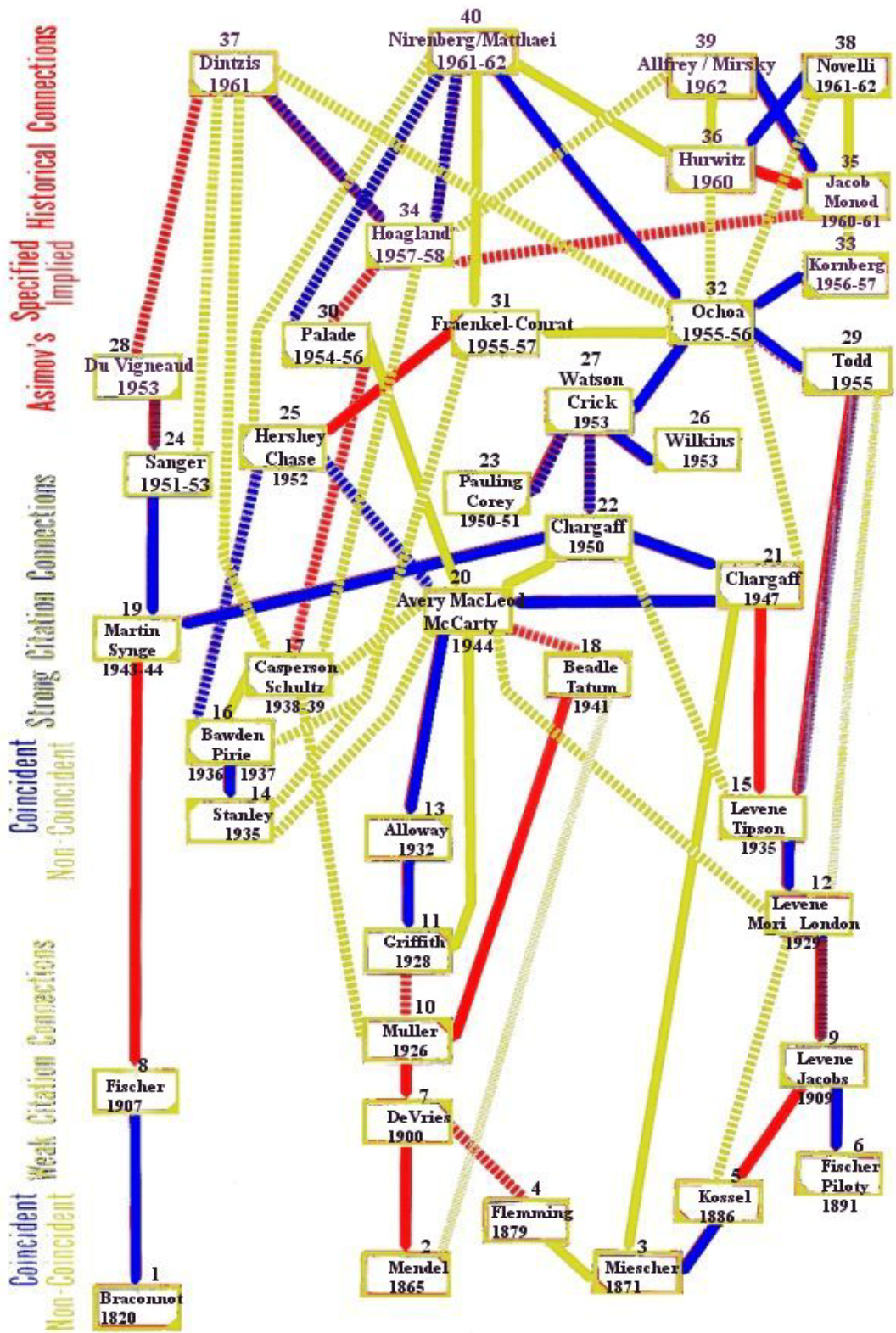

**Figure 1**: Algorithmic Historiogram of the history of DNA from Mendel (1865) to Nirenberg & Matthaei (1961-62). Source: Garfield *et al.*, 1964, at p. 69.

Figure 1 provides the historically first "historiogram" of 1964. The reader will recognize this hand-drawn map as similar to the figures nowadays generated automatically by HistCite™. In his short history, Garfield noted that thereafter he had put the subject to rest for a period of 36 years.

At the time, the development of the field of computational linguistics focused on machine translation and natural language processing. However, the latter problems are technically to be solved at specific moments in time and are therefore static. Citation analysis was in the meantime organized in the *Science Citation Index* on an annual basis, which allows for comparative statics, but not yet for analyzing the dynamics (Leydesdorff, 1991). Hummon & Doreian's (1989) paper triggered a renewed interest in the subject when these authors used citations to analyse main paths in networks over time. Garfield *et al*. (2003a) then returned to the topic of the mapping of the history of DNA using HistCite™, and ever since this program has become a convenient tool for generating algorithmic historiographies which can be combined in various ways with other forms of bibliometric analysis (e.g., Lucio-Arias & Leydesdorff, 2008).

In the different context of science and technology studies, Callon *et al*. (1983) proposed co-word analysis as a means of mapping the dynamics of science. However, the emphasis in this research tradition has been more on transversal "translations" than on measuring longitudinal developments (Callon *et al*., 1986). Kranakis & Leydesdorff (1989) tried to combine a historical analysis with co-word analysis of conference papers using a series of the International Teletraffic Conferences 1955-1983. We concluded that the scientometric analysis using co-word analysis enables us to show variations, whereas the intellectual reconstruction by the historian has to draw a selective line through the history. An evolving complex dynamics then has to be reduced linguistically by using one geometrical metaphor or another.

The combination of (comparative) statics at each moment of time with longitudinal analysis—which necessarily selects either on intellectual grounds (by the historian) or in terms of frequently cited papers (by the citation analyst)—remained a problem to be solved until relatively recently. At each moment of time, one can perform multivariate analysis on a complex dataset (e.g., using cluster of factor analysis). One can then also compare solutions for different time periods, but this comparison remains qualitative. For example, it becomes difficult to control the extent to which the dynamics exhibit the development of error generated by methods which reduce complexity in the (auto-correlated!) data at different moments of time. Addressing this problem dynamically assumes solving a system of partial differential equations in which each variable can change in terms of its value, but also in relation to the structural dimensions (eigenvectors) of the system of variables. Usually, one will not be able to solve this problem analytically.

My own solution of the early 1990s was to use entropy statistics, because in this framework one can use the Shannon (1948) formulas for the static decomposition of the entropy and the Kullback-Leibler (1951) divergence for the dynamic analysis of information in one single framework. In other words, information theory provides us with both a statistics and calculus (Bar-Hillel, 1955; Leydesdorff, 1995; Theil, 1972). Furthermore, by using information theory one remains close to the data, that is, without making parametric assumptions. These methods, however, are computationally intensive (Pearl, 1988) and although exact, sometimes difficult to follow for the social scientist. Furthermore, at the time there was no development of software comparable to that available in SPSS or during the 1990s increasingly from the side of social network analysis.

The problem of combining the static analysis of complexity at each moment of time with a dynamic analysis was solved only recently as an extension to multidimensional scaling (Erten *et al*., 2004; Gansner *et al*., 2005). MDS can be used for the visualization, and dynamic MDS can be extended to the animation (Baur & Schank, 2008). Most techniques for dynamic visualizations are based on smoothing the transitions by linear interpolation between static representations in order to optimize the conservation of a mental map (Moody *et al*., 2005; De Nooy *et al*., 2005; Bender-deMoll & McFarland, 2006). In other words, they are based on comparative statics. The new algorithm, however, allows for optimizing the stress both within each year and over consecutive years, that is, by optimizing the resulting stress in an array of three dimensions or, in other words, a set of stacked matrices.

This new algorithm was implemented in *Visone* (Leydesdorff & Schank, 2008). *Visone* is a software package for the visualization of network data (available at http://visone.info; Baur *et al*., 2002; Brandes & Wagner, 2004). The special version of *Visone* with the dynamic routine added can be web-started from

http://www.visone.info/dynamic/jaws/visone.jnlp or downloaded as stand-alone at

http://www.leydesdorff.net/visone/index.htm. In this study, this program is used for

making dynamic co-word maps of the papers of Eugene Garfield insofar as these have

been included in the citation indices of Thomson Reuters since 1952. I experimented with

this technique for a *Festschrift* on the occasion of the 65th birthday of Michel Callon

(Leydesdorff, 2010) and then found that the maps and animations became more informative when in addition to co-words, co-authors and journal names were used.

Co-words indicate intellectual organization, albeit loosely (Leydesdorff, 1997); co-

authors provide us with social structure, yet without sufficient information about content

(Leydesdorff *et al*., 2008); and the names of journals provide convenient anchor points for structural comparison. As in the previous study, I use cosine-based vector spaces,

five-year time intervals, and only words, authors, and journals that occur more than once

in each specific five-year period. The resulting animation can be retrieved at

http://www.leydesdorff.net/garfield/animation.[1]

**Methods and materials**

Data was collected at the Web of Science on March 15, 2010, for the full period of 1950-

2009, in batches of five-year time periods. (Garfield brings his complete bibliography

online at http://www.garfield.library.upenn.edu/pub.html. This latter data includes

---

[1] The animation of the œuvre of Michel Callon can be retrieved at
http://www.leydesdorff.net/callon/animation .

keynote addresses and publications not included in the (*Social*) *Science Citation Index*.)
Using the citation indexes, 1,546 publications with Garfield as an author were included
until December 31, 2009. Of these documents, 1,080 were published in *Current Contents*
during the period 1973-1993 when the ISI database covered this journal, and 466 were
not (Table 1).

| | Total WoS | Current Contents | This study |
|---|---|---|---|
| 1950-1954 | 3 | 0 | **3** |
| 1955-1959 | 5 | 0 | **5** |
| 1960-1964 | 10 | 0 | **10** |
| 1965-1969 | 46 | 0 | **46** |
| 1970-1974 | 237 | 140 | **97** |
| 1975-1979 | 296 | 269 | **27** |
| 1980-1984 | 294 | 268 | **26** |
| 1985-1989 | 358 | 266 | **92** |
| 1990-1994 | 201 | 137 | **64** |
| 1995-1990 | 48 | 0 | **48** |
| 2000-2004 | 37 | 0 | **37** |
| 2005-2009 | 11 | 0 | **11** |
| Total | 1546 | 1080 | **466** |

**Table 1**: Number of papers of Eugene Garfield under study for each five-year period.

Garfield's papers provide 85.0% of the total number of papers published in *Current Contents* from 1973-1993, when he was also the editor of this journal. The titles of these papers are sometimes exceptionally long; for example: "From Information Scientist to Science Critic – An Introduction to the Role of Information Scientists by Garfield, Eugene, (Reprinted from *The Scientist*, Vol 1, Iss 22, Pg 9, 1987) and Science Needs Critics by Garfield, Eugene, (Reprinted From *The Scientist*, Vol 1, Iss 4, Pg 9, 1987)," *Current Contents* 36, 4 Sep. 1989, 3-7. Because such a title might completely dominate the co-occurrence maps of title words, we decided for this reconstruction to use only the 466 other papers published between 1952 and 2010.

An additional criterion for the selection was that a paper should contain words or coauthor names which occurred at least twice in a five-year period. Using this restriction, 305 papers were eventually included in the exercise. Among the title words of these papers, stop words were removed using the 429 words listed at http://www.lextek.com/manuals/onix/stopwords1.html. Thereafter matrices were constructed for each five-year time period with documents as the units of analysis (the rows) and title words, co-authors, and journal names as variables (in the columns). In order to anchor the visualization, "Garfield" as an author was added as a constant to each case. Among all these variables, cosines can be computed; the threshold for the visualization was set at cosine $\geq 0.2$.

The cosine-normalized matrices were sequenced in a time series using Pajek[2] and dedicated software. The resulting dynamic Pajek project file (with the extension .paj) can be read by Visone, which is then used to generate the animation (see above). Within the animation authors are indicated as circles in red, words in green, and journals as diamonds in blue. Stability is set at the maximum rate of four years. This means that—if available—four periods (of five years) are taken into account for minimizing the stress. (One can compare this with a moving average of four periods, but then multivariately and algorithmically in relation to the stress in the visualization in each period.) The animation results were textually enriched using BlueBerry Flashback (at

[http://www.bbsoftware.co.uk/bbflashback.aspx](http://www.bbsoftware.co.uk/bbflashback.aspx)), a screen-capturing program that allows for editing and the exportation of the results as an Adobe flash or Windows media file.

**Results**

Let me discuss the development sequentially in terms of decades by providing two graphs for each decade.

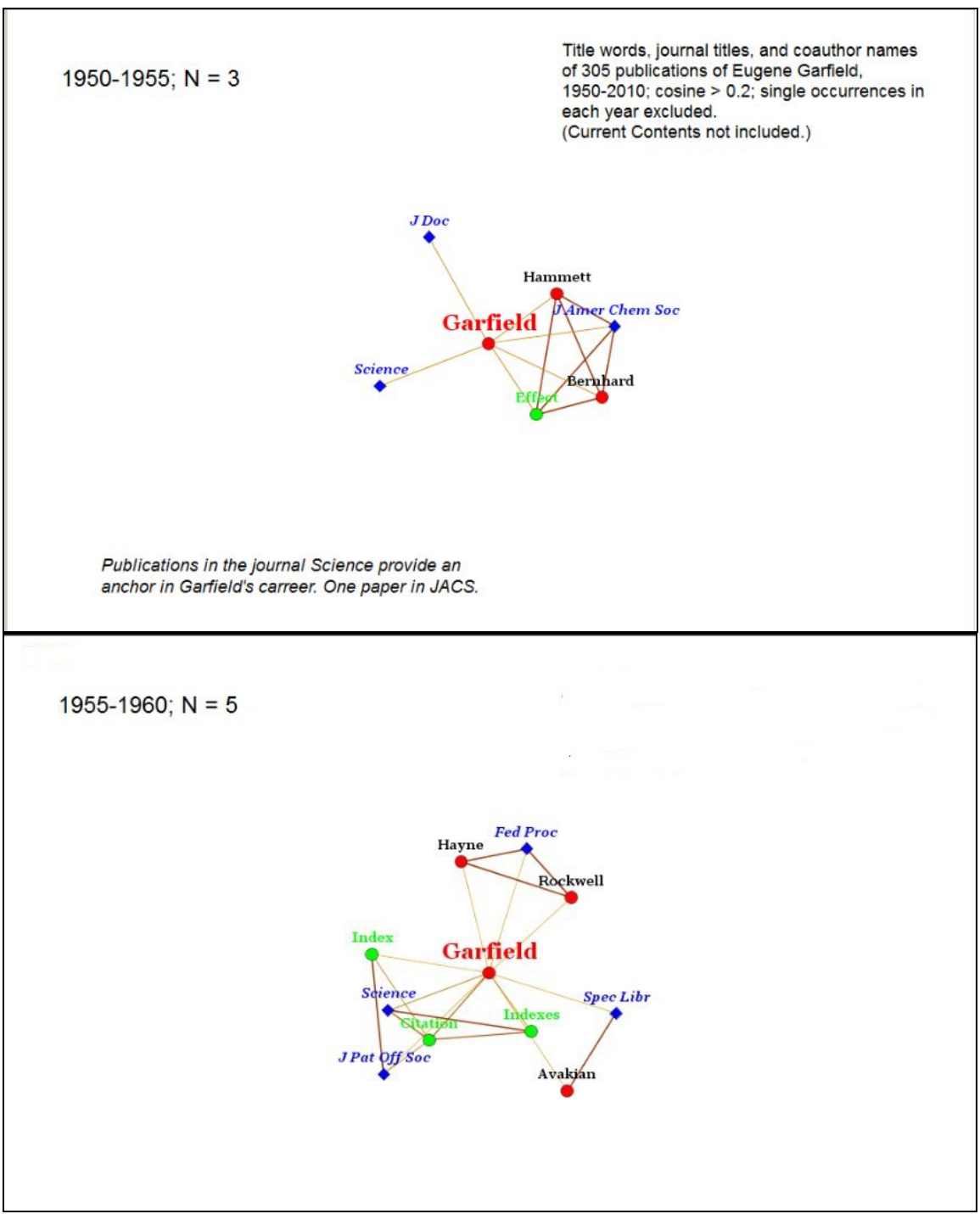

**Figure 2**: The invention of the citation index during the 1950s.

In the first half of the 1950s, Garfield (as a junior scientist) coauthored a publication in the *Journal of American Chemical Society* in 1954 (Thackray & Brock, 2000). In this same year, he published a letter in *Science* and a paper in the *Journal of Documentation*. Probably, the advantages of publishing in *Science* then became clear to him, since as we will see, this journal remains a constant factor during Garfield's further career (Wouters, 1999, 2000).

The period thereafter (1955-1960) witnesses the birth of the citation index as an invention. The eventual innovation—that is, its introduction onto the market—followed much later, with the founding of the ISI and the subsequent organization of the *Science Citation Index* (SCI). The SCI was experimentally published only in 1961. The idea, however, is formulated in Garfield's (1955) paper in *Science* entitled "Citation Indexes for Science: A New Dimension in Documentation through Association of Ideas" (*Science*, 122(3159), pp. 108-111, July 1955). In this article, Garfield proposed to organize scientific citations using the model of *Shephard's Citations*, which has functioned as an index in the legal domain since 1873 (Adair, 1955).

At another place, Garfield (1979a) told the story about the lobbying and persuading that he had to do in order to get the *Science Citation Index* organized (Cronin & Atkins, 2000; Wouters, 1999). The existing patent system, with its standard routines of attributing references to patents both by the applicants and the examiners, may have provided another source of inspiration (Garfield, 1957).

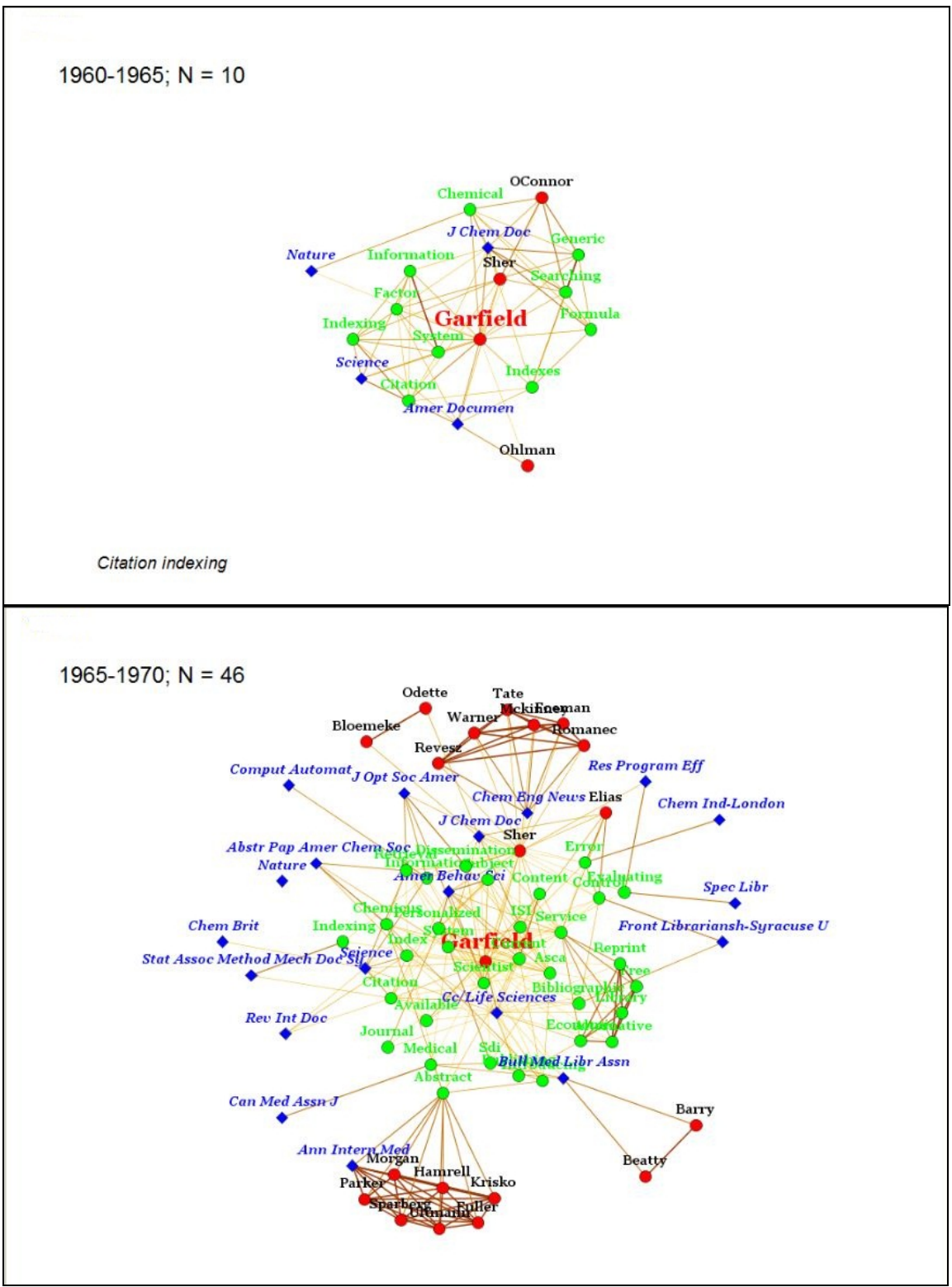

**Figure 3**: Development during the 1960s. Garfield uses his semantics in different directions in co-authorship relations with other groups. The pattern of publications in many journals is established in the second period (1965-1970).

The 1960s, as noted, witnessed the birth of the *Science Citation Index*. Derek de Solla Price (1965) enthusiastically tells the story of how he was invited to play with this encompassing database. which enabled him to write informed history of science at the macro level. Many new ideas were developed in this early period (1960-1965), among them Garfield's already mentioned 1964 report with Sher in which algorithmic historiography was first proposed (Garfield *et al*., 1964). This can be considered as part of a series of studies with Sher in a research line which had to be abandoned thereafter in favour of focusing on the development and diffusion of the new instrument in the period 1965-1975.

In the second period depicted in Figure 3 (1965-1970), one can see the emergence of new vocabularies relevant for the diffusion of the *Science Citation Index*. Garfield still searches co-authorship relations with authors and groups of authors, but new words and additional journals begin to prevail in the representation. The groundwork was laid, for example, in an article in *Science* in 1964, entitled "*Science Citation Index* – A New Dimension in Indexing": the word was now to be spread!

**Figure 4**: The development of semantics and relations with new journals in the 1970s.

Figure 4a shows further developments in the period 1970-1975. A web of words is further shaped without much structure—when compared with the word structure as visible in the animation of Callon's oeuvre (available at

http://www.leydesdorff.net/callon/animation/index.htm)—but relating publications in a wide variety of journals. Co-authors have become more distanced in this vector space. Once established, the relations with journals tend to become more important than the networks of words in the periods after 1975. Perhaps, we may conclude that the semantics of the *Science Citation Index* were shaped in the period 1965-1975, initially with the help of co-authors, but increasingly by Garfield himself. The 1972 article in *Science* entitled "Citation analysis as a tool in journal evaluation" laid the foundation for the use of citations for the evaluation of impact (cf. Garfield & Sher, 1963).

The number of publications is 46 in the period 1965-1970 and 97 during 1970-1975, as against only 26 in the period thereafter (1975-1980). After 1975 co-authorship relations became rare (for example, with Henry Small as a continuous factor in Garfield's environment; e.g., Garfield *et al*., 1978), and the words become even more dispersed so that their occurrence disappears below the thresholds set in this analysis. Relations with journals become increasingly prominent. In the period 1980-1985, co-authorship relations have become nearly invisible, and journal names dominate in Figure 5a. Note the stable relation with the journal *Science* in the various periods. Publications in *Science* functioned as an anchor in Garfield's oeuvre.

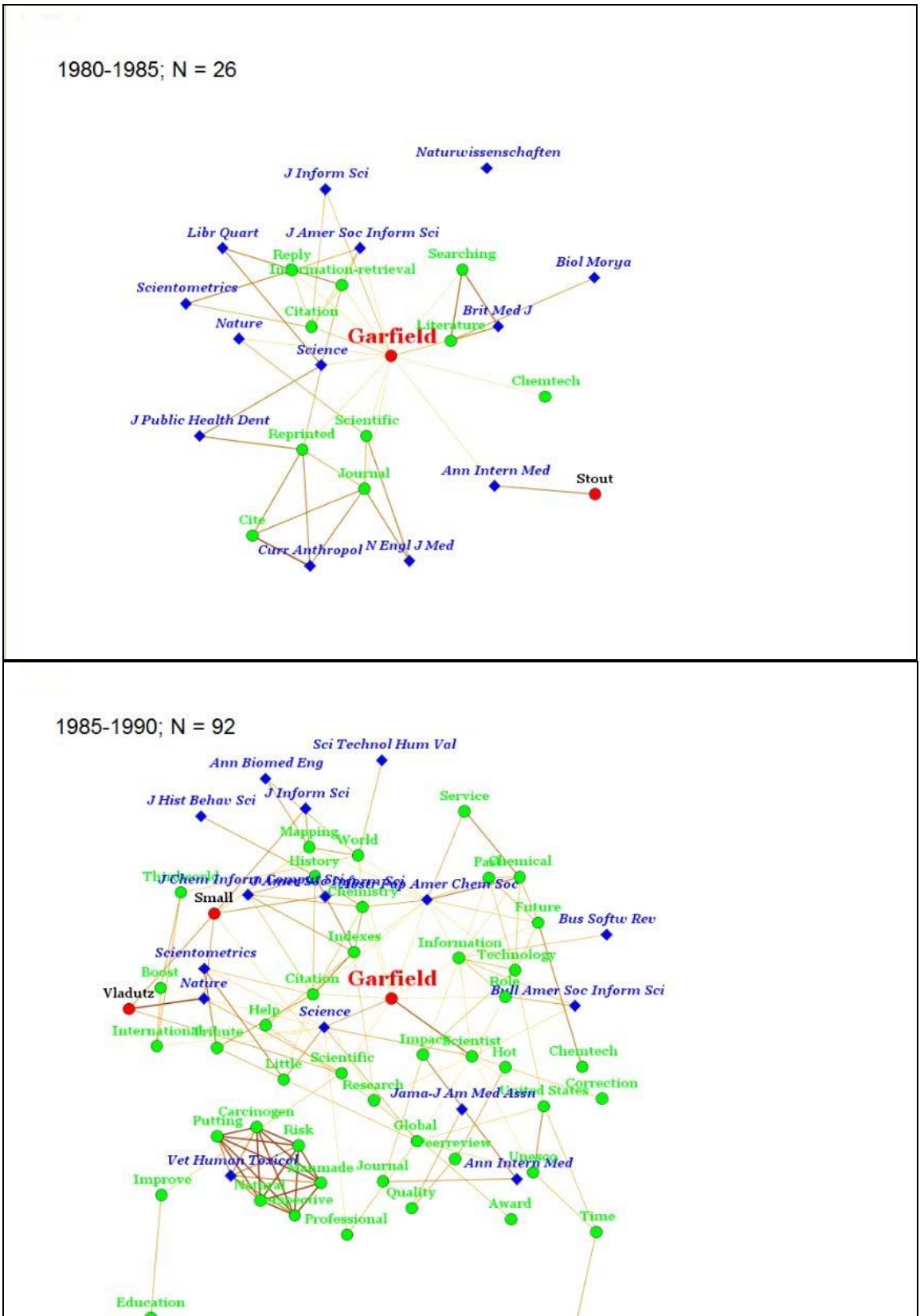

**Figure 5**: The 1980s: new applications of citation analysis.

In the latter half of the 1980s, the intellectual program was renewed with papers in a large number of journals, including many in the information sciences. In this period, citation indexing is reconsidered no longer as only a tool for retrieval, but advocated as a research instrument (e.g., Garfield, 1978, 1979b; Small & Garfield, 1985; cf. Leydesdorff, 1987). New terminology is organized for publications in the newly established journal of *Scientometrics* (1978) and in *Science, Technology, and Human Values*, which in 1987 became the journal of the *Society for the Social Studies of Science*. (Garfield's Institute of Scientific Information has supported the yearly Bernal Prizes of this organization.) The turn to the information sciences, however, prevails in Figure 5b.

**Figure 6**: Garfield as an information scientist in the 1990s.

As in previous decades, publications in *Science* are central to the development of Garfield's oeuvre during the 1990s. This journal seems "close to his heart" in terms of the development of patterns of collaboration, semantics, and publications. The turn to the information sciences is now unequivocal. This will eventually be acknowledged by his election as President of the *American Society for Information Science* during the period 1998-2000. During his presidency, Garfield proposed renaming the society and its journal as *American Society for Information Science and Technology*: *JASIS* accordingly became *JASIST* in 2001. In the second half of this period (1995-2000) the network begins to shrink as a result of his reaching retirement age (Cronin & Atkins, 2000).

**Figure 7**: The period 2000-2010: retirement and renewed interest in algorithmic historiography (2000-2005).

In the first half of the decade 2000-2010, co-authorship relations re-enter the scene. For example, and of most relevance to our topic, the program HistCite™ is developed in collaboration with Alexander Pudovkin, and a series of publications follows. Other relations with leading scientometricians and information scientists are also visible during this period. In the final period (2005-2010), the relations with journals—perhaps on invitation—again seem most prominent. The network now shrinks and the number of publication decreases to eleven.

**Conclusions and discussion**

Algorithmic bibliography can enrich the reconstruction—beyond the drawing of a historiogram showing citation linkages—with words, co-authorship relations, journal names, or any variable which can be attributed to a communication as a unit of analysis. Note that these units of analysis can also be aggregated into oeuvres (or document sets) which can be compared with one another using these same techniques. Like the historiography, citation linkages highlight specific lineages across the document sets (Kranakis & Leydesdorff, 1989). The analysis then presumes a selective perspective either reflexively, by the qualitative analyst, or methodologically, by the citation analyst who chooses, for example, to focus on highly-cited papers.

Other tools have been developed in bibliometrics: citation analysis was extended to co-citation analysis invented almost simultaneously by Small (1973) and Marshakova (1973); bibliometric coupling as the reverse operation preceded this invention (Kessler, 1963); co-word analysis was developed by Callon *et al.*, (1983), and journal mapping by Leydesdorff (1986, 1987) and Tijssen *et al.* (1987), etc. All these structural properties can be mapped on top of one another as overlays (Leydesdorff *et al.*, 2008). A consensus has grown in the community that Salton's cosine can be used for spanning a vector space in which these vectors can be positioned (Ahlgren *et al.*, 2003; Egghe & Leydesdorff, 2009; Salton & McGill, 1983) and then mapped using a spring-embedded algorithm (e.g., Kamada & Kawai, 1989) or multi-dimensional scaling in one form or another (Van Eck & Waltman, in print).

In addition to the intellectual lineages shown by (co-)citation analysis, co-authorship relations may enable the analyst to show elements of social structures (Burt, 1983; Persson, 2004; Wagner, 2008; Zitt *et al.*, 2000). Co-word analysis can add substantive content to network representations and thus facilitate reading. Journal names, for example, add symbolic value, so that the readers can orient themselves in terms of where to locate these markers of intellectual organization (Leydesdorff, 1989, 1997). In Leydesdorff (2010), the static map was stepwise decomposed into co-words, co-authors, and journal names in order to show how the addition of layers of heteronegeous networking can add up to the construction of a more informed representation.

The focus on lineage in citation analysis and historiography can with hindsight be understood as an attempt to reduce complexity when both the variables and the structures are subject to change (Leydesdorff, 1997, 2002). Entropy statistics allows us to perform multivariate analysis dynamically (Leydesdorff, 1991, 1995), but the visualization and *a* *fortiori* the dynamic animation had failed us hitherto for the multivariate case. This was recently solved using a new algorithm that optimizes the majorant and thus reduces stress in the multidimensional scaling including the time dimension. Formulated at a more abstract level, one can thus animate any variable that can be attributed to texts; for example, one can show the shifting knowledge bases of (groups of) authors by using the journal names in their cited references and their bibliographic coupling with BibJourn.Exe (available at http://www.leydesdorff.net/software/bibjourn/index.htm). Thus, shifts in the knowledge bases of nations can be mapped and animated, in principle. A new domain of knowledge visualization and animation can thus be made a subject of research.

In this study, I wished to demonstrate a specific application of this technique on the occasion of Eugene Garfield's 85[th] birthday, using his own works as recorded in the *Science* and *Social Science Citation Indexes.* The intellectual history thus written is different from a personal history (Thackray & Brock, 2000) or an institutional history (Wouters, 1999, 2000) of the same events. Furthermore, we excluded the hundreds of contributions to *Current Contents* because the (sometimes very long) titles of these articles would lead us astray from our purpose of portraying the intellectual development. I did not focus on citations among the attributes because one can conveniently obtain the citation graph using HistCite™. The present analysis adds to a historiogram (e.g., Figure 1) the multivariate perspective. The dynamic version of *Visone* was specifically developed with this purpose in mind (Leydesdorff & Schank, 2008).

**Acknowledgement**

I am grateful to Katy Börner for her (hitherto informal) collaboration on this subject.